\documentstyle[aps,preprint,prd,eqsecnum,epsfig,tighten,floats]{revtex}
\def\bn{{\bf n}}

\def\bP{{\bf P}}
\def\bx{{\bf x}}

\def\bsigma{\bbox{\sigma}}
\def\Tr{\mathop{\text{Tr}}}
\def\muhat{{\hat\mu}}
\def\nuhat{{\hat\nu}}

\begin{document}
\title{Topology and metastability in the lattice Skyrme model}
\author{Alec J. Schramm}
\address{Department of Physics, Occidental College, Los Angeles,
California 90041}
\author{Benjamin Svetitsky}
\address{School of Physics and Astronomy, Raymond and Beverly Sackler
Faculty of Exact Sciences, \\
Tel Aviv University, 69978 Tel Aviv, Israel}
\date{\today}
\maketitle
\begin{abstract}
We offer the Skyrme model on a lattice as an effective field theory---fully
quantized---of baryon--meson interactions at temperatures below the chiral
phase transition.
We define a local topological density that involves the volumes of tetrahedra
in the target space $S^3$ and we make use of Coxeter's formula for
the Schl\"afli function to implement it.
This permits us to calculate the mean-square radius of a skyrmion in the
three-dimensional lattice Skyrme model, which may be viewed as a Ginzburg-Landau
effective theory for the full quantum theory at finite temperature.
We find that, contrary to expectations, the skyrmion shrinks as quantum and
thermal fluctuations are enhanced.
We ascribe this to a large number of metastable states that become accessible
as the temperature is raised.
\end{abstract}
\section{Introduction}
The Skyrme model \cite{Skyrme,Witten,ANW,Bal,Zahed} is a popular model of
the dynamics of pions and nucleons, incorporating the former as its fundamental,
pseudo-Goldstone field and the latter as topological solitons.
The continuum theory has been widely studied via semiclassical techniques,
giving a satisfactory phenomenology of low-momentum and low-temperature
physics.
In this paper we develop a lattice formulation of the model.\footnote{For
previous work see \cite{Saly,DeTar}.}

Our motivation in employing a lattice cutoff
is to overcome the limitations of the continuum theory.
In point of fact, the term ``continuum theory'' is misleading.
The Skyrme Lagrangian is a non-linear sigma model with a four-derivative
term, which makes it non-renormalizable in perturbation theory.
This means that calculations of quantum effects must involve a short-distance
cutoff.
In working with chiral Lagrangians coupled to a local nucleon field, this
is not a serious problem \cite{Weinberg,GL,Bijnens}.
One absorbs divergences into an ever-lengthening list of counterterms;
as long as external momenta are kept small,
dimensional analysis limits the contributions of these counterterms to the
results.
When the nucleon is a soliton rather than a fundamental field, however,
things are more difficult.
The soliton's collective degrees of freedom are quantized separately, and
the systematic development of higher-order quantum effects involves
disentangling the pion field from these collective quanta \cite{Mattis}.
In any case, there is no way to limit the addition of higher-dimension
terms to the Lagrangian as long as one must consider energy scales
approaching the nucleon mass.

We propose to turn these points to advantage, by considering the Skyrme
model as an effective field theory.
The cutoff, and the cutoff scheme (the lattice), are part and parcel of
the specification of the theory.
The free choice of terms in the Lagrangian is now a virtue, constrained only
by phenomenological necessity.
The procedure will be to pick a Lagrangian; to fix the lattice spacing;
and finally to calculate any physical quantity of interest, going beyond
perturbation theory and beyond semiclassical methods, since the lattice
offers many more direct techniques.
From a lattice point of view, the fixed cutoff is a virtue as well,
since no continuum limit is necessary (nor is it possible).

The most interesting feature of the Skyrme model is the stability
conferred on the soliton by the topological conservation law and
the four-derivative coupling.
We wish in fact to isolate a single skyrmion in order to see how its
properties change with temperature, density, etc.
A recurring problem in lattice studies of solitons is the tendency
of lattice dislocations to destroy topological stability or, in other
words, to allow topology to slip through the lattice.
We avoid this problem by choosing a lattice action that enforces
continuity at short distances.
In order to keep our numerical evolution within a single topological sector, 
we insist on a local updating scheme that should not nucleate smooth
skyrmions.
We find that these two ingredients suffice to stabilize the lattice
skyrmion.

The main technical development in this paper is the construction of an
exact topological density that can be measured on a lattice field
configuration.
By ``exact'' we mean that the winding number, the sum of the density over
the lattice, is always an integer.
We cut the lattice into fundamental tetrahedra and map each tetrahedron
into a curved tetrahedron in the $S^3$ target space.
The winding number is then the sum of the signed volumes of the tetrahedra in
$S^3$.
We calculate the volume of a spherical tetrahedron via a formula
due to Coxeter \cite{Coxeter}, derived as a solution of differential
equations first written down by Schl\"afli \cite{Schlafli}.

We summarize the continuum Skyrme model in Sec.~II in order to establish
notation.
General considerations regarding continuity and topology on the lattice
are presented in Sec.~III, which concludes with formulae for the lattice
action we employ.
In Sec.~IV we present our definition of the local topological density
via Coxeter's formula for the volume of a quadrirectangular tetrahedron
in $S^3$.
As a first application, we present in Sec.~V the results of Monte Carlo
simulation for the classical Skyrme model in three dimensions.
Easier to simulate than the full four-dimensional path integral, this
model may be regarded as a Ginzburg-Landau theory for the Skyrme model
at finite temperature---a sort of effective theory for the effective theory.
We use the topological density to calculate the mean-square radius of
a skyrmion as a function of the couplings $\beta_1$ and $\beta_2$ of the
lattice action.
If we fix $\beta_2/\beta_1$, so that the form of the action is fixed,
we find a multitude of metastable configurations of the skyrmion that are
smaller than the ground state solution.
As the ``temperature'' is raised by decreasing $\beta_1$ and $\beta_2$, these
metastable states are made accessible to fluctuations, with the result that
the skyrmion shrinks as it is ``heated.''
Presumably the ``temperature'' of the 3d theory is an increasing function
of the real temperature in the 4d theory that it approximates.
Thus we reach the result that a quantized skyrmion shrinks as it is heated.

\section{Continuum Skyrme model}
The Euclidean action of the Skyrme model is
\begin{equation}
S=\int d^4x \left[\frac{f_\pi^2}{16}\Tr|\partial_\mu U|^2
+\frac1{32e^2}\Tr([L_\mu,L_\nu])^2\right],
\label{continuum_action}
\end{equation}
where the non-linear chiral field $U$ is an $SU(2)$ matrix, 
and we have defined
\begin{equation}
L_\mu=iU^{\dag}\partial_\mu U.
\end{equation}
We have omitted a mass term of the form $m_\pi^2\Tr U$ and thus
$S$ is invariant under the $SU(2)\times SU(2)$
group of chiral rotations,
\begin{equation}
U\to AUB^{\dag},\text{ with }A,B\in SU(2).
\end{equation}

Classically, the symmetry is spontaneously broken.
Finite-energy field configurations must tend to a constant at infinity, and one
can use a symmetry rotation to make this constant the unit matrix, i.e.,
\begin{equation}
U(\bx)\to \openone\text{ as } |\bx|\to\infty.
\label{Uinfty}
\end{equation}
Using the Pauli matrices $\tau_i$, $i=1,2,3$, we can write $U$ in terms of
new fields $\bsigma=(\sigma_0,\sigma_i)$ via
\begin{equation}
U=\sigma_0+i\sigma_i\tau_i,
\label{Usigma}
\end{equation}
where $\bsigma\cdot\bsigma=1$.
By considering small fluctuations about $U=\openone$,
we can identify $\sigma_i$, $i=1,2,3$ with the Goldstone pion field.

Equation~(\ref{Uinfty}) means that 3-dimensional space is compactified to
the 3-sphere $S^3$.
Since $U\in SU(2)$ also takes values in $S^3$ [see Eq.~(\ref{Usigma})],
configurations $U(\bx)$ can be classified according to the homotopy group
$\pi_3(S^3)=Z$.
Thus there is an integer winding number
$n$ that denotes how many times $U(\bx)$ covers the 3-sphere in
field space as \bx\ is varied over its 3-sphere.
This winding number is topologically conserved, meaning that it cannot change
under continuous deformation of the field $U(\bx)$.

The $n=0$ sector includes the vacuum $U(\bx)=\openone$ and perturbations around
it.
An example of an $n=1$ configuration is the spherically symmetric
skyrmion,
\begin{equation}
U(\bx)=\exp\left[if(r)\frac{x_i\tau_i}r\right],
\label{skyrmion}
\end{equation}
with
\begin{eqnarray}
f(0)&=&\pi,\nonumber\\
f(\infty)&=&0.
\end{eqnarray}
$f(r)$ should be determined so as to minimize the static energy
\begin{equation}
E=\int d\bx \left[\frac{f_\pi^2}{16}\Tr|\partial_i U|^2
+\frac1{32e^2}\Tr([L_i,L_j])^2\right],
\label{continuum_E}
\end{equation}
but a simple choice with the right topology is
\begin{equation}
f(r)=\pi\left(1-\tanh\frac r{r_0}\right).
\label{tanh}
\end{equation}
Skyrme identified $n$ with the baryon number of a field configuration,
and the lowest soliton configuration with the nucleon.

Given an arbitrary field configuration $U(\bx)$ satisfying Eq.~(\ref{Uinfty}),
its winding number may be calculated with the formula
\begin{equation}
n=\frac{i}{24\pi^2}\int d\bx\,\epsilon_{ijk}\Tr L_iL_jL_k.
\label{continuum_n}
\end{equation}
The geometric meaning of Eq.~(\ref{continuum_n}) will be apparent in its
lattice counterpart below.

\section{Lattice topology and continuity}

In this section we define a topological density for a lattice field
configuration that is unambiguous and that sums exactly to an integer.
The further demand of conservation of the winding number will lead
us to choosing a lattice action that constrains discontinuities in
the field.

\begin{figure}[ht]
\centerline{\psfig{figure=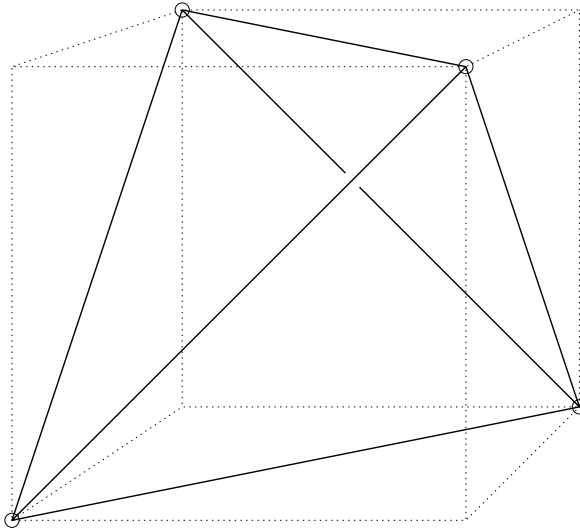,height=7cm}}
\caption{A cube cut into 5 tetrahedra.  The even vertices (circled)
are connected to form a central tetrahedron and four others.
In neighboring cubes, the {\em odd\/} vertices are to be connected instead,
so that the diagonal edges match up.}
\label{tetra5}
\end{figure}

A lattice configuration is specified by the field $U_\bn\in SU(2)$ or,
equivalently, by the 4-vector $\bsigma_\bn\in S^3$.
In order to define the winding number, we begin \cite{DeTar}
by cutting the cubic
lattice $L^3$ into tetrahedra, five tetrahedra per cubic cell
(see Fig.~\ref{tetra5}).
The four vertices $\bn^{(i)}$, $i=1$--4 of each tetrahedron map to four
unit 4-vectors $\bsigma^{(i)}$, which are vertices of a spherical
tetrahedron in $S^3$.
This tetrahedron is defined via its vertices; its edges are arcs of
great circles and its faces are spherical triangles drawn on great spheres.

Since three (non-collinear) points in $S^3$ determine a great sphere, it
is clear that two adjacent tetrahedra in $L^3$ map into adjacent tetrahedra
in $S^3$, with the common face in $L^3$ mapping into a common face in $S^3$.
Thus the field configuration $\bsigma_\bn$ gives a triangulation of some
volume in $S^3$.
We impose periodic boundary conditions on the lattice, and thus no tetrahedron
possesses a face that is not shared by another tetrahedron (a {\em free\/}
face).
This implies that the complex of tetrahedra in $S^3$ possesses no free faces
either, and thus the volume covered by the complex must be an integer multiple
of the volume\footnote{
The three-dimensional volume of the unit 3-sphere is $2\pi^2$.
We will redefine this to be unity, and thus measure volumes in units
of $2\pi^2$.}
of $S^3$.
If the field configuration is smooth, the complex in $S^3$ will be composed
of tetrahedra that are small compared to $S^3$, and it will wrap around
the sphere much as a smooth mapping of the 3-torus must.
In any case, the volume in $S^3$ of the five tetrahedra corresponding to
a cube in $L^3$ gives a definition of the topological density $\rho_\bn$ 
contained in the cube.

The topological density thus defined is not unambiguous, however.
If a tetrahedron in $S^3$ is specified by its faces, then there are two
volumes in $S^3$ that are bounded by these faces.
One of the volumes includes the north pole (for example), and the other
does not.
If one of the volumes is measured to be a positive $V$ (with $V<1$), then
the other volume will be $V-1$.
(Careful attention to the orientation will make the latter negative.)
In order to assign a unique topological density
to a lattice field configuration,
we define a tetrahedron's volume to satisfy $|V|<\frac12$.

We have solved the problem of uniqueness, but not that of conservation of
the winding number.
Consider a field configuration wherein one tetrahedron in $S^3$ has volume
$\frac12-\epsilon$, where $\epsilon$ is small.
Under a fluctuation of one of the vertices of the tetrahedron, its volume may
shift to $\frac12+\delta$, while the neighboring tetrahedra change their
volumes by $-(\epsilon+\delta)$ so that the winding number is unchanged.
Unfortunately, our algorithm will now redefine the volume of the first
tetrahedron to $-\frac12+\delta$, resulting in a loss of 1 in the winding
number.
This is often called ``topology dropping through the lattice.''

The procedure to be followed at this point depends on the physics to be
investigated.
One might want to study, for example, the thermodynamics of the Skyrme
model by fixing a chemical potential $\mu$ coupled to the winding number.
The grand partition function would then be
\begin{equation}
Z(\mu)=\sum_n e^{\mu n} Z_n,
\end{equation}
where $Z_n$ is a sum over all field configurations with winding number $n$.
A Monte Carlo simulation should then be allowed to wander freely among the
sectors of different $n$, subject to an acceptance/rejection test that
enforces the relative probabilities $e^{\mu n}$.
A local updating algorithm can nucleate ``point skyrmions'' as in the preceding
paragraph, which change $n$ by a unit and then spread out into smoother
skyrmions;
in addition,
a non-local update could be permitted that creates smooth skyrmions directly.

Our interest, however, is in the properties of a single skyrmion,
which are accessible through a canonical ensemble at fixed $n=1$.
The simulation must be constrained so as not to change $n$.
This requires that we prevent the nucleation of both point skyrmions and
smooth skyrmions.
The latter can be prevented by choosing a local updating algorithm.
For the former, we choose a lattice action that excludes the possibility
of tetrahedra in $S^3$ with $|V|\approx\frac12$.
We adopt the kinetic term \cite{Ward}
\begin{equation}
S_1=(\alpha-1)\sum_{\bn\mu}\log(\bsigma_\bn\cdot\bsigma_{\bn+\muhat}-\alpha)
\label{S1}
\end{equation}
in order to constrain $\bsigma_\bn\cdot\bsigma_{\bn+\muhat}>\alpha$ and thus 
to reject
fluctuations that put large angles between neighboring field variables.
With some Monte Carlo exploration, we can find a value of $\alpha$ that
will keep the tetrahedral volumes far enough from $\frac12$.

For smooth configurations, we expand $\bsigma_\bn\cdot\bsigma_{\bn+\muhat}$
around 1 and Eq.~(\ref{S1}) becomes
\begin{eqnarray}
S_1&\simeq&\sum_{\bn\mu}(1-\bsigma_\bn\cdot\bsigma_{\bn+\muhat})\nonumber\\
&=&\sum_{\bn\mu}\left[1-\frac12 \text{Tr}\,\left(U_\bn U_{\bn+\muhat}^{\dag}
\right)\right]\nonumber\\
&=&\sum_{\bn\mu}\frac14\text{Tr}\,\left(U_{\bn+\muhat}-U_\bn\right)
\left(U_{\bn+\muhat}^{\dag}-U_\bn^{\dag}\right),
\end{eqnarray}
which approaches the kinetic term in Eq.~(\ref{continuum_action}) in the
continuum limit.

A technical point remains.
We have discussed the ambiguity in fixing the volume of a tetrahedron
if its faces are given, which led us to require $|V|<\frac12$.
A lattice field configuration $\bsigma_\bn$ gives us only the vertices of
each tetrahedron, however, not its faces.
Given three vertices that determine a great sphere, the face that connects
them can be chosen to be either of two triangles that together make up the
sphere.
A little thought shows that the difference between the volumes enclosed
is $\frac12$.\
This gives an ambiguity between a value $V>0$ and $V-\frac12$ for the volume
of the tetrahedron.

This last ambiguity involves choosing which triangle constitutes the face
of the tetrahedron.
But this face is shared between two adjacent tetrahedra, and the ambiguity
will be immaterial if the same choice is made for both.
This is hard to program, however, if each tetrahedron is to be handled on
its own.
We prefer to resolve the ambiguity for each tetrahedron separately, by
requiring that $|V|<\frac14$.
We impose this via a more restrictive choice of $\alpha$ in Eq.~(\ref{S1}).
We find that setting $\alpha=0.1$ is adequate for the purpose.

A local updating algorithm based on the action (\ref{S1}) will conserve
winding number.
Even so, a skyrmion will tend to shrink down to a point in accordance
with the scaling argument that gives Derrick's Theorem.
This is because the stabilization is being done at the scale of a single
lattice spacing;
only here will the scaling argument fail and the skyrmion run into a
repulsive potential.
In order to have a stable skyrmion of appreciable size, we cannot avoid
adding a Skyrme term as in Eq.~(\ref{continuum_action}).
The most straightforward lattice transcription uses symmetric derivatives
\cite{Saly},
\begin{equation}
S_2^{\text{SYM}}=\sum_\bn\sum_{\mu>\nu}\left\{
\left(\bsigma_{\bn+\muhat}-\bsigma_{\bn-\muhat}\right)^2
\left(\bsigma_{\bn+\nuhat}-\bsigma_{\bn-\nuhat}\right)^2-
\left[
\left(\bsigma_{\bn+\muhat}-\bsigma_{\bn-\muhat}\right)\cdot
\left(\bsigma_{\bn+\nuhat}-\bsigma_{\bn-\nuhat}\right)
\right]^2\right\}.
\end{equation}
This couples each site to sites two links away, effectively via plaquette
couplings across plaquettes of side $\sqrt2$, and fails to couple the odd
and even sublattices.
We prefer to use a discretization \cite{Adler} that couples
only across single plaquettes, of side 1:
\begin{equation}
S_2=4\sum_\bn\sum_{\mu>\nu}\left\{
\left(\bsigma_{\bn+\muhat}-\bsigma_{\bn+\nuhat}\right)^2
\left(\bsigma_{\bn+\muhat+\nuhat}-\bsigma_{\bn}\right)^2-
\left[
\left(\bsigma_{\bn+\muhat}-\bsigma_{\bn+\nuhat}\right)\cdot
\left(\bsigma_{\bn+\muhat+\nuhat}-\bsigma_{\bn}\right)
\right]^2\right\}.
\end{equation}
Our lattice action is a combination of the kinetic and Skyrme terms,
\begin{equation}
S=\beta_1S_1+\beta_2S_2.
\label{action}
\end{equation}
In the naive continuum limit, we can compare with Eq.~(\ref{continuum_action})
to deduce that
\begin{equation}
\beta_1=\frac{f_\pi^2}4,\qquad
\beta_2=\frac1{32e^2}.
\end{equation}

\section{Topological density on the lattice}

As shown in Fig.~\ref{tetra5}, we cut each cube on the lattice into five
tetrahedra.
Each tetrahedron maps into a curved tetrahedron in $S^3$, the volume of
which gives the local topological density.
We give here a practical, exact formula for this volume.

In the 2-sphere $S^2$, the area of a triangle is given simply by Girard's
theorem, $A=\alpha+\beta+\gamma-\pi$, where the right hand side is the sum
of the triangle's angles minus $\pi$, known as the angular excess.
The easiest proof of the theorem is an argument based on overlapping lunes.
This argument may be generalized \cite{Sommerville} to a simplex in
$S^n$, but it only gives a formula for the volume when $n$ is even;
for odd $n$, one obtains instead a constraint on the angles of the simplex.

\begin{figure}[ht]
\centerline{\psfig{figure=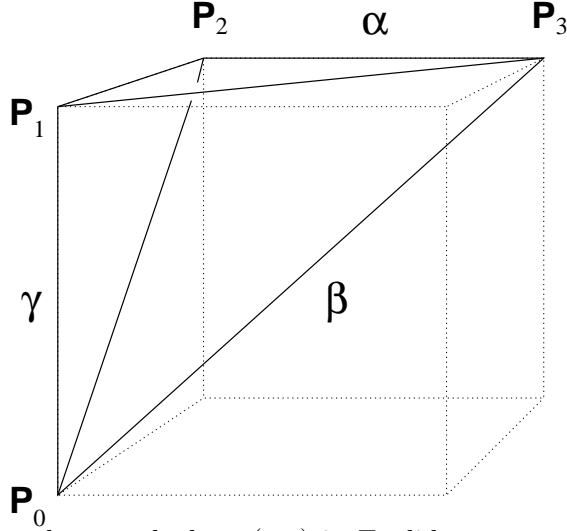,height=7cm}}
\caption{A quadrirectangular tetrahedron (qrt) in Euclidean space.
The path from $\bP_0$ to $\bP_3$ connects opposite vertices of a rectangular
solid.}
\label{qrtetra}
\end{figure}

L.~Schl\"afli \cite{Schlafli} attacked the general problem of the volume
of convex polytopes in $S^n$.
He was led to consider (for $n=3$) the special case of the
{\em quadrirectangular tetrahedron\/}, which we shall abbreviate as {\em qrt.}
A qrt (see Fig.~\ref{qrtetra}) is constructed as follows \cite{Coxeter}.
Choose four points $\bP_0$, $\bP_1$, $\bP_2$, $\bP_3$
such that the line segments
$\bP_0\bP_1$, $\bP_1\bP_2$, $\bP_2\bP_3$
are all mutually perpendicular.
Then connect all four points together to form a tetrahedron.
Every face of this tetrahedron is a right triangle.
Three of the dihedral angles---the ones at $\bP_1\bP_2$,
$\bP_1\bP_3$, and $\bP_0\bP_2$---are right angles.
The dihedral angles at
$\bP_2\bP_3$, $\bP_0\bP_3$, and $\bP_0\bP_1$
are {\em not} right angles, and they are denoted
$\alpha$, $\beta$, $\gamma$.

This construction works equally well in Euclidean space and in $S^3$.
A simple example of a spherical qrt is constructed by taking
$\bP_0$, $\bP_1$, $\bP_2$, $\bP_3$
to be mutually perpendicular unit 4-vectors.
This is a qrt that covers 1/16 of $S^3$.
Its dihedral angles are all $\pi/2$.

In Euclidean space, the dihedral angles of a qrt satisfy
\begin{equation}
\sin\alpha\,\sin\gamma=\cos\beta.   \label{Euclidean}
\end{equation}
In $S^3$, they satisfy instead the inequality
\begin{equation}
\sin^2\alpha\,\sin^2\gamma>\cos^2\beta.   \label{spherical}
\end{equation}
The quantity
\begin{equation}
D=\sqrt{\sin^2\alpha\sin^2\gamma-\cos^2\beta}
\end{equation}
may thus be taken to be the generalization of the angular excess to this case.
It vanishes as the qrt becomes small, which is the Euclidean limit.

Let us denote the volume of a spherical qrt as $V(\alpha,\beta,\gamma)$,
normalized  to $2\pi^2$ for the entire 3-sphere.
Schl\"afli derived formulas for the derivatives $\partial V/\partial\alpha$,
$\partial V/\partial\beta$, $\partial V/\partial\gamma$
(see also \cite{Richmond}).
The integral of these formulas was found by Coxeter \cite{Coxeter} in the form
of a Fourier series valid in the restricted domain
\begin{eqnarray}
0\leq&\alpha&\leq\frac12\pi\nonumber\\
0\leq&\beta&\leq\pi\nonumber\\
0\leq&\gamma&\leq\frac12\pi.
\label{domain}
\end{eqnarray}
Expressing the volume in terms of the complements of $\alpha$ and $\gamma$,
\begin{equation}
V(\alpha,\beta,\gamma)=\frac14S
\left(\frac\pi2-\alpha,\beta,\frac\pi2-\gamma\right),
\end{equation}
his solution for the Schl\"afli function $S$ is
\begin{equation}
S(x,y,z)
= \sum_{m=1}^\infty\left(\frac{D-\sin x\sin z}{D+\sin x\sin z}\right)^m
\frac{\cos 2mx-\cos 2my+\cos 2mz-1}{m^2}
-x^2+y^2-z^2.
\label{Schlafli}
\end{equation}

\begin{figure}[ht]
\centerline{\psfig{figure=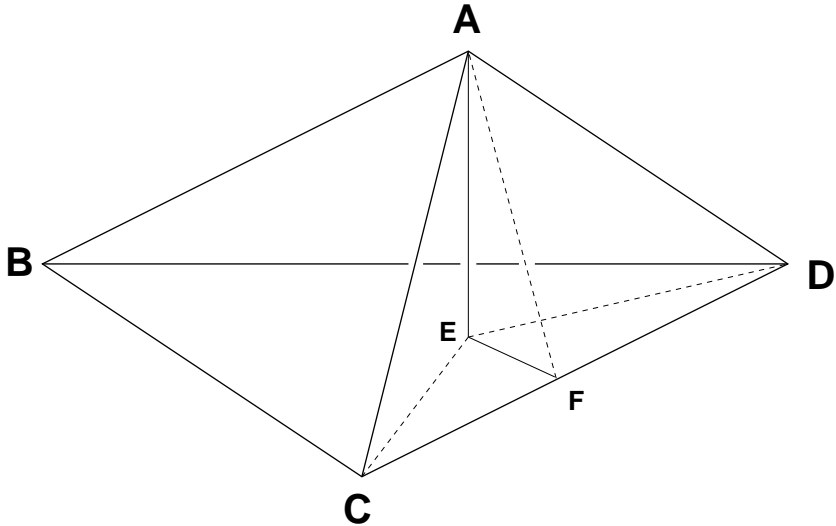,height=7cm}}
\caption{A Euclidean tetrahedron cut into six qrt's.
(Only two are shown.)}
\label{cut_tetra}
\end{figure}

Armed with Eq.~(\ref{Schlafli}), we can calculate the volume of any
tetrahedron in $S^3$ by cutting it up into six qrt's (see Fig.~\ref{cut_tetra}).
This is done by picking a vertex {\bf A} of the tetrahedron and dropping
a perpendicular to the opposite face (the base) at {\bf E}, then dropping
perpendiculars from {\bf E} to the edges of the base.
Simple trigonometry gives the vertices of the qrt's in terms of those of
the original tetrahedron; from the vertices of each qrt we calculate its
dihedral angles $\alpha, \beta, \gamma$.
Some technical problems, as well as further discussion of the Schl\"afli
function, are relegated to the Appendix.

The topological density $\rho_\bn$ in the lattice cube at \bn\ is the sum of the
volumes of the tetrahedra in $S^3$ that correspond to the Euclidean tetrahedra
making up the cube.
It goes without saying that summing $\rho_\bn$ over the lattice must give an
integer (in units of $2\pi^2$)
for any configuration, to high precision.

\section{Classical Skyrme model in 3 dimensions}

The minima of the action (\ref{action}) in 3 dimensions are the static,
classical solutions of the 4-dimensional Skyrme model.
When we include fluctuations about these minima, we can think of the 3d
action that governs them as of an approximation to an effective action
derived via dimensional reduction.
Thus the parameters $\beta_1,\beta_2$ are (unknown) functions of the parameters
of the full 4d theory, namely, $f_\pi,e$, and the temperature $T$.
Naturally, any reduction scheme that is precisely defined will give an
effective action that is far more complex than (\ref{action}), and further
arguments will be necessary to justify its simplification.

As an application of our algorithm for the topological density we calculate
the mean-square radius $R^2$ of a single skyrmion in the three-dimensional
theory, that is, of an equilibrium distribution of
configurations with $n=1$ at fixed $\beta_1,\beta_2$.
We fix initial conditions of the form (\ref{skyrmion}) and (\ref{tanh}),
suitably discretized.
Monte Carlo updates are done with the usual local Metropolis algorithm, which
preserves winding number as discussed above.
The volume of the lattice is $16^3$.

The observable $R^2$ should be defined carefully.
For each configuration used in the average one might calculate
the barycenter of the
topological density according to
\begin{equation}
{\bf R}_c=\sum_\bn \bn\rho_\bn, 
\label{Rc}
\end{equation}
and then the second moment via
\begin{equation}
R^2=\sum_\bn \text{min}(\bn-{\bf R}_c)^2 \rho_\bn .
\label{R2}
\end{equation}
(The notation ``min'' means that one should take note of the periodic
boundary conditions and always calculate the shortest distance between
\bn\ and ${\bf R}_c$.)
Unfortunately, the boundary conditions make
${\bf R}_c$ calculated via Eq.~(\ref{Rc}) ill-defined.
This may be illustrated by considering a compact skyrmion with barycenter
located on the face of the lattice at $z=0$.
Equation~(\ref{Rc}) fixes ${\bf R}_c$ in this case
to be in the middle of the lattice, between
the two half-skyrmions on opposite faces, and $R^2$ will turn out to
be on the order of the lattice size.
The solution lies in regarding the lattice as a 3-torus on which the choice of
origin (and of ``faces'') is entirely arbitrary.
We evaluate Eqs.~(\ref{Rc}) and (\ref{R2}) for all choices of origin and
take the minimal value of $R^2$.
The ${\bf R}_c$ that corresponds to this minimum is thus defined to be
the location of the barycenter;
this is as good a definition as any, considering the periodic boundary
conditions.

\begin{figure}[ht]
\centerline{\psfig{figure=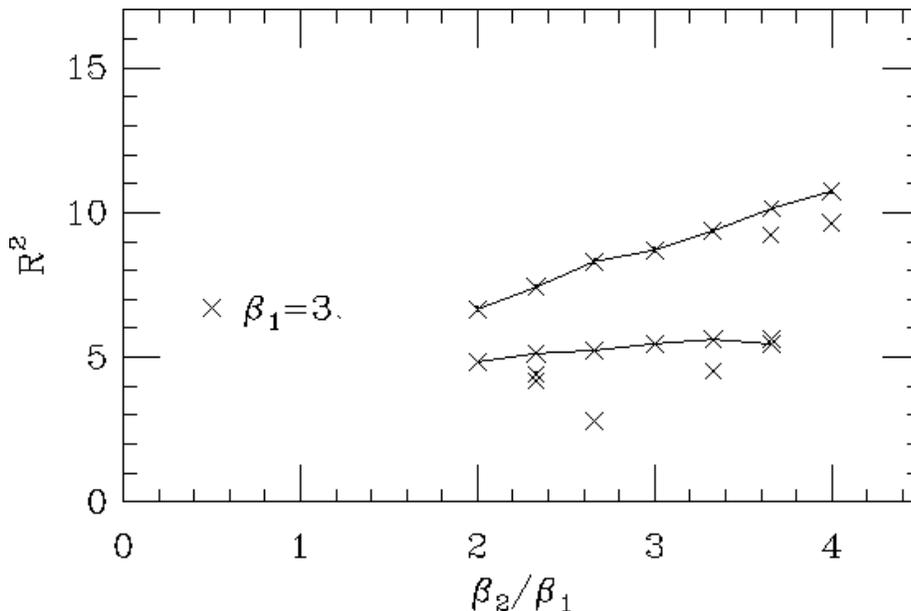,height=8cm}}
\caption{Mean-square radius of a single skyrmion, $\beta_1=3$.
The upper curve connects the radii of stable configurations;
other points are radii of metastable skyrmion configurations.}
\label{r2b3}
\end{figure}

We show in Fig.~\ref{r2b3} the calculated skyrmion radius for $\beta_1=3$
and various values of $\beta_2$.
The multiple values of $R^2$ at a given value of $\beta_2$ reflect 
{\em metastability}.
These metastable skyrmion configurations are accessible from different
initial states, specified by different values of $r_0$ in Eq.~(\ref{tanh}).
We have observed tunneling from metastable states to lower-energy states
of larger radius that are apparently the global minima of the action.
In all cases the lowest-energy states appear to be those of largest radius;
these are shown connected by the upper curve in Fig.~\ref{r2b3}.
The lower curve in the figure connects metastable states that were found
to be mutually accessible by annealing.
We have {\em not\/} explored the full set of metastable states for any given
coupling;
except for the annealing curve, the metastable states shown in Fig.~\ref{r2b3}
are effectively chosen at random by the initial conditions of the
simulations.

\begin{figure}[ht]
\centerline{\psfig{figure=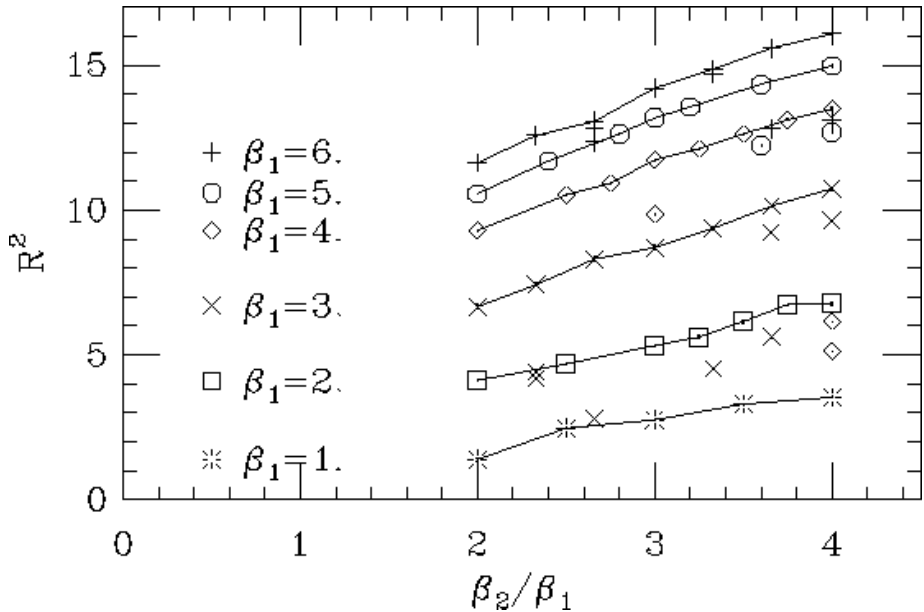,height=8cm}}
\caption{Mean-square radius of the topological density in the single-skyrmion
sector}
\label{r2}
\end{figure}

In Fig.~\ref{r2} we present a more extensive set of results, this time
showing variation with $\beta_1$ as well as with $\beta_2$.
The curves connect data points for what we believe to be the equilibrium
skyrmion configuration at each coupling.
Points not lying on curves are metastable configurations of higher energy;
again, we make no effort to show all such configurations at any coupling.
We have not run across metastable configurations for $\beta_1\leq 2$.

The abscissa in Fig.~\ref{r2} is the ratio $\beta_2/\beta_1$ that fixes
the form of the action.
The remaining overall coefficient $\beta_1$ then acts as an inverse temperature
that governs the fluctuations.
Regarding the 3d classical model as a Ginzburg-Landau theory for the fully
quantized 4d theory, the lowering of $\beta_1$ can be interpreted as
enhancing the thermal and quantum fluctuations in the latter.
It is remarkable that heating the skyrmion by lowering $\beta_1$ causes it
to {\em shrink}.
We ascribe this surprising property to the many metastable, excited
states that are
smaller than the ground-state configuration and that become accessible
as the temperature is raised.
The absence of metastability at the smallest values of $\beta_1$ merely
reflects the fact that the equilibrium skyrmion already averages over
the various local minima.

\section{Discussion}


The qualitative conclusion of the preceding section must be taken with a large
grain of salt.
As mentioned above, we have no idea what the connection is between
the couplings $\beta_1$ and $\beta_2$ of the 3d model on the one hand and
the couplings and temperature of the 4d theory on the other.
It is possible that raising the physical temperature of the latter will
lead to simultaneous changes in $\beta_1$ and $\beta_2$ that cause the
skyrmion to expand after all.
In any case, the result is not quantitative and cannot be made quantitative
until the dimensional reduction to the 3d theory is explored in detail.
The metastable solutions, for that matter, might be artifacts
of the specific lattice action we study that are absent in the true
effective 3d theory.

Our study of the 3d model was motivated by its
simplicity rather than by any fundamental obstacle to studying the full
4d theory at finite temperature.
Numerical work on the 4d theory will merely require greater computer
resources.
In order to extract physics from lattice calculations, we will have to
renormalize the theory by fixing the bare couplings $\beta_1$ and
$\beta_2$, and the lattice spacing $a$,
in terms of physical quantities such as the $\pi\pi$ scattering length
and the skyrmion mass and radius.
The temperature can be varied by changing the time extent $L_t$ of the
lattice in the time direction, or by varying the lattice spacing by changing
$\beta_1$ and $\beta_2$ along lines of constant physics.
The 4d quantum theory will offer as well the possibility of projecting
the skyrmion to definite spin and isospin.
Finally, it will be imperative to study the sensitivity of any physical
quantities to the choice of lattice action.

\section*{Acknowledgments}
We thank M.~Kugler and U.-J.~Wiese for their help.
Conversations with the latter took place at the Aspen Center for Physics.
We also thank the Weizmann Institute of Science for its hospitality.
A.~J.~S. thanks the Theoretical Elementary Particle Physics
group at UCLA for its hospitality during his sabbatical leave.
The work of B.~S. was supported in part by the Israel Science Foundation
under Grant No.~255/96-1.
The work of A.J.S. was supported by the Research Corporation.
\appendix
\section*{}
We collect in this appendix some practical details concerning the use of the
Schl\"afli function (\ref{Schlafli}) in calculating the topological density
in lattice field configurations.

\subsection{Small qrt's}

Defining
\begin{equation}
X\equiv\frac{\sin x\sin z-D}{\sin x\sin z+D},
\end{equation}
the formula for the Schl\"afli function is
\begin{equation}
S(x,y,z)
= \sum_{m=1}^\infty (-X)^m \frac{\cos 2mx-\cos 2my+\cos 2mz-1}{m^2}
-x^2+y^2-z^2.
\label{Schlafli2}
\end{equation}
The sum converges absolutely when $|X|\leq1$; since $D>0$, this means
$\sin x\sin z>0$, which includes the entire domain (\ref{domain})
considered by Coxeter.
In the Euclidean limit, where the qrt is small,
we have $D\to0$ and hence $X\to1$.
Then the sum in Eq.~(\ref{Schlafli2}) becomes the Fourier series for
$x^2-y^2+z^2$, and thus $S=0$ in this limit as expected.

The region of small $D$ is where the series converges most slowly.
We can avoid evaluating Eq.~(\ref{Schlafli2}) in this region as follows
\cite{DeTar}.
The vertices of a small tetrahedron in the 3-sphere are marked by 4-vectors
$\bsigma^{(i)}$ that are nearly equal.
These 4-vectors are edges of a spindly 4-pyramid with its apex at the center
of the 4-ball bounded by the 3-sphere, the volume of which is
\begin{equation}
V_4=\frac1{24}\epsilon_{\mu\nu\rho\sigma}\sigma^{(1)}_\mu \sigma^{(2)}_\nu
\sigma^{(3)}_\rho \sigma^{(4)}_\sigma.
\end{equation}
The ratio of $V_4$ to the total volume $\pi^2/2$ of the 4-ball is
approximately the ratio of the 3-volume $V$ of the tetrahedron to the total
volume $2\pi^2$ of the 3-sphere, so
\begin{equation}
V\approx4V_4.
\label{VV4}
\end{equation}
We calculate $V_4$ for every tetrahedron obtained for each lattice
cube (see Fig.~\ref{tetra5}).
If it is small, we use Eq.~(\ref{VV4}) in lieu of cutting it into qrt's.
We also calculate $V_4$ for each qrt in turn, and if it small we use Eq.~(\ref{VV4})
in lieu of evaluating the Schl\"afli function.
Evaluation of $V_4$ also gives an easy way to keep track of the sign of
a qrt's volume, which is needed since $S(x,y,z)$ is defined always to be
positive.

\subsection{Large tetrahedra}

Another problem region for the Schl\"afli function arises when
$\alpha\to\frac\pi2$ or 
$\gamma\to\frac\pi2$, giving $X\to-1$ and slow convergence.
In the limit $X=-1$, we have
\begin{equation}
S(x,y,z)=\pi(-x+y-z)=\pi\left(\alpha+\beta+\gamma-\pi\right),
\end{equation}
a useful approximate formula in this regime.

\begin{figure}[ht]
\centerline{\psfig{figure=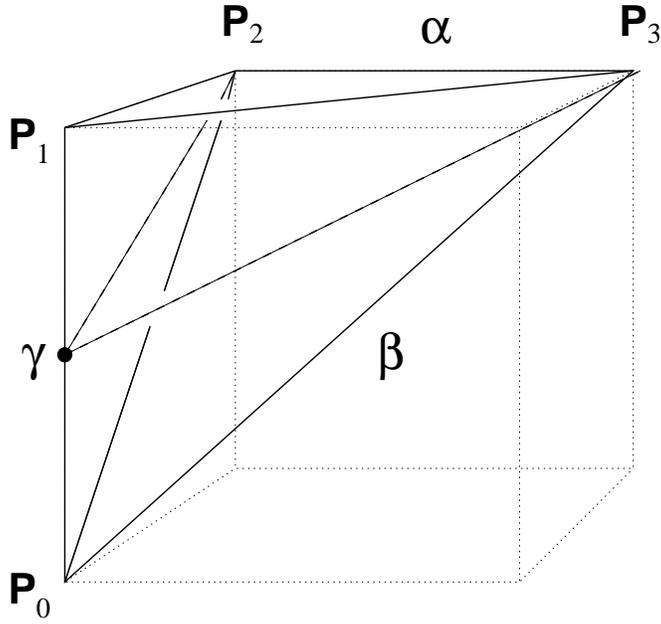,height=8cm}}
\caption{Cutting a qrt in the event that $\alpha$ is too large.}
\label{cut_qrt}
\end{figure}

As we have noted, Coxeter's formula can only be used in the domain
(\ref{domain}).
One can easily devise larger qrt's, however, and large qrt's arise regularly
in calculating the topological density.
We attack this problem by cutting such large qrt's into smaller tetrahedra.
If either $\alpha$ or $\gamma$ is larger than $\frac\pi2$, but not both,
we determine the largest of $\alpha,\beta,\gamma$ and draw a plane through its
edge and the bisection point of the opposite edge (see Fig.~\ref{cut_qrt}).
This creates two new, smaller tetrahedra.
If both $\alpha$ and $\gamma$ are larger than $\frac\pi2$, we cut the
tetrahedron by connecting all its edge centers; this gives four tetrahedra
at the original vertices, plus a central octahedron.
The octahedron can then be cut into four tetrahedra around any line
connecting opposite vertices, of which we choose the shortest.

Having cut the original qrt into two or eight pieces, we apply the algorithm
(cut into qrt's and apply the Schl\"afli function) to each piece.
Of course, any qrt that results might still be too large for Coxeter's formula.
We then apply the cutting algorithm recursively.
We have found that ten levels of recursion might be necessary for particularly
rough lattice configurations.

The amount of recursion can be cut down dramatically by optimizing the
operation of cutting the original tetrahedron into qrt's.
There are four ways to choose the vertex {\bf A} in Fig.~\ref{cut_tetra}.
For each choice of {\bf A}, we determine the longest edge of
the generated qrt's, measured by the largest pairwise angle among the 4-vectors
$\bsigma^{(i)}$ of each qrt.
This maximal edge is a measure of the size of the qrt's generated by the
cutting process.
We choose {\bf A} in order to minimize the maximal edge.

\end{document}